\def\fgw{f_{\rm GW}}
\def\mdens{{\rm g~cm^{-3}}}
\def\bdens{{\rm fm^{-3}}}
\def\msol{M_{\odot}}

\documentclass[times]{aa}
\usepackage{graphicx}

\begin{document}

\title{Impact of the nuclear equation of state on the last orbits
of binary neutron stars}

\author{ M. Bejger\inst{1}, D. Gondek-Rosi\'{n}ska\inst{1,2},
E. Gourgoulhon\inst{2}, P. Haensel\inst{1,2},
K. Taniguchi\inst{3}\fnmsep\thanks{\emph{Present address:}
Department of Physics,
University of Illinois at Urbana-Champaign,
Urbana, Illinois 61801, USA}, 
J. L. Zdunik\inst{1} }
\institute{
N. Copernicus Astronomical Center, Polish Academy of
Sciences, Bartycka 18, PL-00-716 Warszawa, Poland
\and
Laboratoire de l'Univers et de ses Th\'eories,
UMR 8102 du C.N.R.S., Observatoire de Paris, 
Universit\'e Paris 7, F-92195 Meudon Cedex,
France
\and
Department of Earth Science and Astronomy,
Graduate School of Arts and Sciences,
University of Tokyo, Komaba, Meguro, Tokyo 153-8902, Japan
}
\offprints{M. Bejger, \email{bejger@camk.edu.pl}}
\date{Received xxx Accepted xxx}
\abstract{We present calculations of quasiequilibrium sequences
of irrotational binary neutron stars based on
realistic equations of state (EOS) for the whole neutron star
interior. Three realistic nuclear EOSs of various softness and
based on different microscopic models have been joined with a recent
realistic EOS of the crust, giving in this way three different
EOSs of neutron-star interior. Computations of quasiequilibrium
sequences are performed within the Isenberg-Wilson-Mathews
approximation to general relativity. 
For all evolutionary sequences,
the innermost stable circular orbit (ISCO) is found to be given 
by mass-shedding limit (Roche lobe overflow). 
The EOS dependence on the last orbits is found to be quite important:
for two $1.35~M_{\odot}$ neutron stars,
the gravitational wave frequency at the ISCO (which
marks the end of the inspiral phase) ranges from 800~Hz to 
1200~Hz, depending upon the EOS. Detailed comparisons with 3rd order
post-Newtonian results for point-mass binaries reveals a very 
good agreement until hydrodynamical effects (dominated by high-order 
functions of
frequency) become important, 
which occurs at a frequency ranging from 500~Hz to 1050~Hz, 
depending upon the EOS.
 }

\titlerunning{Nuclear EOS and the last orbits of BNS}
\authorrunning{Bejger et al. }
\maketitle
%
\section{Introduction}
\label{intro}
%
Detection of the gravitational waves (GW) with the use of the
ground-based laser interferometers, such as VIRGO (\cite{VIRGO}),
LIGO (\cite{LIGO}), GEO600 (\cite{GEO}), or TAMA300 (\cite{TAMA})
will provide crucial information about
various astrophysical objects. Among them, coalescing binary
neutron stars (NS) seem particularly interesting in order to probe
their interiors. Matched filtering techniques involving high-order
post-Newtonian effects will provide the individual masses and spins of the
NSs. The last few orbits before the final merger are dominated by
the strong tidal forces acting between the components. The tidal
deformation of NSs shape and of the matter distribution in the
stellar interior are expected to depend rather strongly on the
poorly known equation of state (EOS) of dense matter.  The GW
signal carries therefore some imprint of the EOS.  In particular,
the final frequency of the last stable circular orbit is strongly
correlated with the compactness parameter $M/R$ (\cite{Faber02,
BNSvar03}), and thus can provide direct constraints on the theory
of dense hadronic matter  above the nuclear saturation density. 
A related event potentially rich of informations about the dense matter
EOS is the merger of neutron star and a black hole
(\cite{PrakaL04,PrakaRL04}). 
In addition to  locating the transition from inspiral to the merger
phase, computations of the last stable orbits of binary NSs are
also required for providing initial data to compute the dynamical
merger phase (\cite{ShibaTU03} and references therein).

The last orbits of inspiraling binary neutron stars have been
studied by a number of authors in the quasiequilibrium
approximation, and in the framework of Isenberg-Wilson-Mathews
(IWM) approximation of general relativity (see \cite{BaumgS03} for
a review). The quasiequilibrium assumption approximates the
evolution of the system by a sequence of exactly circular orbits
(as the time evolution of the orbit is much larger than the
orbital period). The IWM approximation amounts to using a
conformally flat spatial metric (the full spacetime metric
remaining non conformally flat), which reduces the problem to
solving only five of the ten Einstein equations. Within these two
approximations, the most realistic studies have considered {\em
irrotational} binaries, as opposed to {\em synchronized} ones, for
the viscosity of neutron star matter is far too low to ensure
synchronization during the late stage of the inspiral
(\cite{BildsC92}, \cite{Kocha92}). 

The quasi-totality of the existing quasiequilibrium IWM studies
(Bonazzola et al. 1999, Maronetti et al. 1999, Uryu \& Eriguchi 2000,
Uryu et al. 2000, Gourgoulhon et al. 2001, Taniguchi \& Gourgoulhon
2002b, 2003) employ
a polytropic EOS to model the neutron star interior. The only
exception is the recent work of \cite{OechsUPT04} who have used
two EOSs : (i) a pure nuclear matter EOS, based on a relativistic
mean field model and (ii) a `hybrid' EOS with a phase transition
to quark matter at high density. At $2\times 10^{14}\ {\rm g\,
cm}^{-3}$ (i.e. in the vicinity of nuclear density), 
these two EOSs are matched to a polytropic one with
adiabatic index $\gamma=2.86$. This last assumption of 
Oechslin et al. is somewhat ad hoc, 
because the EOS of neutron star crust is very
different from a polytrope, and its local adiabatic index is much
smaller. Indeed, the  crust polytropic index within the inner
crust (which contains some 99.9\% of the total crust mass) varies
from $\gamma\simeq 0.5$ near the neutron drip point to 
$\gamma\simeq 1.6$ in the bottom layers near
the crust-core interface (Douchin \& Haensel 2001).
It is to be noticed that the crustal EOS plays a crucial
role in defining the mass-shedding limit which marks the
end point of quasiequilibrium binary configurations. 

In the present article, we study the last orbits of irrotational
binary neutron star systems by using a set  of three dense  matter
EOSs which are representative of the contemporary many-body theory
of dense matter.  Contrary to \cite{OechsUPT04}, we describe the
neutron star crust by means of a realistic  EOS obtained in the
many-body calculations. As in the works mentioned above, we use
the quasiequilibrium and IWM approximations.

The paper is arranged in the following way: in Sect. \ref{EOSdescr}
the differences between EOSs used in the computation are briefly
described. Sect. \ref{method} is devoted to a description of the
numerical methods used to obtain the quasiequilibrium orbital sequences.
In Sect. \ref{results} the results are presented, whereas Sect.
\ref{conclusion} contains discussion and final remarks.

%
\section{Description of the equations of state}
\label{EOSdescr}
%
\subsection{The EOS of the crust}
\label{EOScrust}
The outer layer of neutron star contains neutron rich nuclei,
which due to Coulomb repulsion form a crystal lattice if matter
temperature is below the corresponding melting temperature. This
layer is called neutron star crust, and extends up to the density
at the crust-core interface, $\rho_{\rm cc} \sim 10^{14}~\mdens$.
The precise value of $\rho_{\rm cc}$ is model dependent, and
varies within $(0.6-1.4)\times 10^{14}~\mdens$. The EOS of the
crust is rather well established (for review, see, e.g.,
\cite{Haens03}). In the present paper the EOS of the crust was
composed of three segments. For densities smaller  than
$10^{8}~\mdens$ we used the EOS of \cite{BPS}. For $10^8~\mdens
<\rho<10^{11}~\mdens$ we applied the EOS of \cite{HaensP94}, which
makes maximal use of the experimental masses of neutron rich
nuclei. Finally, for densities $10^{11}~\mdens<\rho<\rho_{\rm cc}$
we used the EOS of \cite{SLy01}. For neutron stars of $M=1.35~{\rm
M}_\odot$, the crust contains less than 2\% of stellar mass.
However, it is the region  most easily deformed under the action
of the tidal forces resulting from the gravitational field
produced by the companion star. Below melting temperature, elastic
shear terms in the stress tensor are nonzero, but they are two
orders of magnitude smaller than the main diagonal pressure term
(\cite{Haens01}). Dissipative and thermal effects accompanying
matter flow inside neutron stars will be briefly discussed in
Sect. \ref{conclusion}; they are expected to be small at  the
quasiequilibrium evolution stage. Therefore, to a very good
approximation, the crust layer behaves in the tidal forces field
as an ideal degenerate fluid, described by a zero temperature EOS.

An important quantity which actually determines the response of
the crust layer to the compression or decompression of matter is
the adiabatic index $\gamma=(n/p){\rm d}p/{\rm d}n$, where $p$ is
the local pressure and $n$ the corresponding  baryon number
density. In the {\it outer crust}, the pressure is determined by the
ultrarelativistic electron gas, so that $\gamma=4/3$. However, the
outer crust contains only $10^{-5}$ of the NS mass. Some thousand
times more, i.e, about $0.03M_\odot$, is contained in the {\it
inner crust}, composed of a lattice of heavy neutron rich nuclei
immersed in an neutron gas and an electron gas. Its adiabatic
index depends strongly on the density, varying from $\gamma\simeq
0.5$ near the neutron drip point (extreme softening of the EOS),
up to $\gamma \simeq 1.6$ close to $\rho_{\rm cc}$ (Douchin \&
Haensel 2001).
\subsection{The EOS of the core}
\label{EOScore}
%
Neutron star core contains matter of density significantly larger
than the normal nuclear density, equal to the saturation density
of infinite symmetric nuclear matter, $\rho_0=2.8\times
10^{14}~\mdens$, and corresponding to the baryon
 number density $n_0=0.16~\bdens$. For $\rho>\rho_0$ the EOS of the core
is poorly known, and this uncertainty grows rapidly with
increasing density. The theoretical EOSs, derived using different
theories of dense matter, and different methods of solution of the
many-body problem, differ significantly at $10^{15}~\mdens$,
characteristic of the central cores of neutron stars under study. 
 In the present paper, we will limit ourselves
to neutron star cores consisting of nucleons and hyperons, i.e.,
baryons whose properties are known from terrestrial experiments.
The speculative case of exotic phases of dense matter (kaon and
pion condensation, quark deconfinement), whose existence is - as
for this writing -  not substantiated neither by experiments nor
by astronomical observations, will be considered in future
publications. The most extreme case of hypothetical strange stars
built of self-bound strange quark matter will also be studied
separately.

Three EOSs of the core were considered. Two of them may be
considered as a soft and stiff  extremes of the EOSs of matter
composed of nucleons, electrons, and muons. The first of them,
BPAL12,  is of phenomenological type and can be considered as a
soft extreme of the nucleonic EOS of NS matter 
(\cite{PrakaBPELK97}, see also \cite{Bombaci95}).
The second one, APR,  is based on precise variational calculations
and includes realistic two-body (Argonne A18) and three-body
(Urbana UIX) nucleon interactions (\cite{akmal})
\footnote{The APR EOS violates the causality at high density region
which appears in NS core only close to 
 the maximum mass of the neutron star.}.
We considered
also one EOS in which hyperons are present 
 at $\rho>\rho_{\rm H}$, where 
the threshold for the hyperon appearance $\rho_{\rm H}\simeq
2\rho_0$ (model 3
of \cite{Glend1985}); it will be referred to as the GNH3 EOS. This
EOS was obtained using the Relativistic Mean Field model of
baryonic matter. For $\rho<\rho_{\rm H}$ (nucleons only) this EOS
is very stiff but causal ($v_{\rm sound}<c$). The appearance of
hyperons strongly softens the EOS as compared to the pure nucleon
case.  The hyperons soften the EOS because they are more massive
than nucleons,  and when they start to fill their Fermi sea they
are slow and replace the highest  energy  nucleons.
\begin{figure}[h]
\centering
\resizebox{3.25in}{!}{\includegraphics[clip]{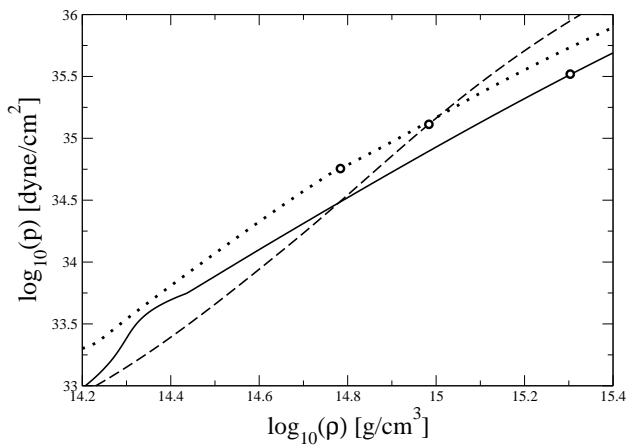}}
\caption{The pressure $p$ against the energy density $\rho$ for
the EOSs used in the paper: APR (dashed line), GNH3 (dotted line),
and  BPAL12 (solid line).  The empty circles correspond to the
central density of a non-rotating stellar model with a
gravitational mass equal to $1.35~M_{\odot}$ (Table~\ref{EOStable})
}
\label{fig:eospvsrho}
\end{figure}


\begin{figure}[h]
\centering \resizebox{3.25in}{!}{\includegraphics[clip]{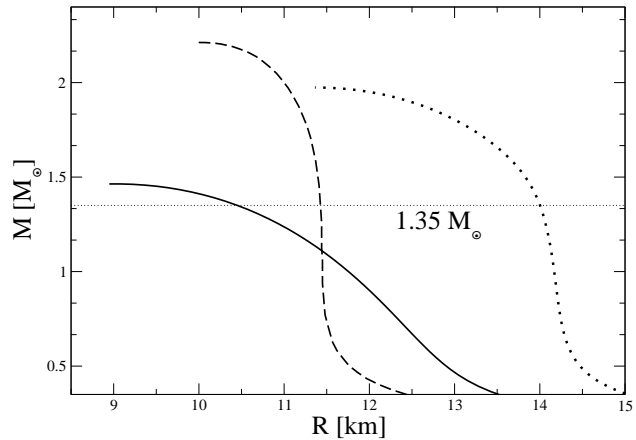}}
\caption{Gravitational mass of static isolated stars against
areal radius for the APR (dashed line), GNH3 (dotted line),
and BPAL12 (solid line) EOSs.} \label{fig:mvsr}
\end{figure}


All three EOSs  are displayed in Fig. \ref{fig:eospvsrho}. As we
mentioned they are very different because they assume very
different strong interactions models  at supranuclear densities.

\subsection{Mass-radius relation for static NSs}
\label{MRcurves}
The differences among the three EOSs 
are reflected in the $M(R)$ curves of
static stable models shown in Fig. \ref{fig:mvsr}
($M$ being the gravitational mass and $R$ the areal radius). 

The BPAL12 EOS is overall quite soft, and therefore produces NSs
with $R$ decreasing rapidly with increasing $M$. The BPAL12 EOS
yields the smallest $R(M)$ for $M>1.2M_\odot$. The  maximum
allowable mass for this EOS is marginally consistent with
observations, $M_{\rm max}=1.46~{\rm M}_\odot$.  The APR EOS is
stiff, with radius staying nearly constant at $R\simeq 11.5~$km
for $M=(0.6-1.6){\rm M}_\odot$, and yields high $M_{\rm max}
=2.2~{\rm M}_\odot$. The BPAL12 and APR EOSs are  soft and stiff
extremes within our set of the EOSs as far as the values of
$M_{\rm max}$ are concerned. However, their  $M(R)$ intersect at
$M=1.2M_\odot$. Therefore, these EOSs produce NSs with similar
values of compaction parameter $M/R$ for $M\simeq 1.2M_\odot$.

The GNH3 EOS is different from the two previous ones, because its
$M(R)$ curve is composed of two different segments. The lower-mass
segment ($M<1.3~{\rm M}_\odot$) consists of stars with no or with
only small hyperon cores. The radius stays nearly constant at
$R\simeq 14$ km for $M=(0.5-1.3){M}_\odot$. This segment is
connected via a ``knee'' with a high-mass segment consisting of
neutron stars with increasingly larger soft hyperon cores. Along
this high-mass (hyperon-softened) segment,  $R$ decreases rapidly
with increasing $M$, reaching a very flat maximum at $M\simeq
1.9~{\rm M}_\odot$.

\begin{table}[h]
\begin{center}
\begin{tabular}{c|c|c|c|c}
\multicolumn{1}{c}{EOS} & \multicolumn{1}{|c|}{$M/R$} &
\multicolumn{1}{|c|}{$R~[{\rm km}]$}&
\multicolumn{1}{|c|}{$M_{\rm B} [M_{\odot}]$}&
\multicolumn{1}{c}{$\rho_c~[10^{14}{\rm g/cm^3}]$} \\
\hline
\hline
GNH3 & 0.140 & 14.262 & 1.45351 & 6.26 \\
\hline
APR & 0.176 & 11.350 & 1.49110 & 9.80 \\
\hline
BPAL12 & 0.191 & 10.447 & 1.48472 & 20.22 \\
\end{tabular}
\end{center}
\caption{Properties of isolated static neutron stars of
gravitational mass $M=1.35\, M_\odot$ for the three EOSs used in
our computations. $M/R\equiv GM/Rc^2$ is the compaction parameter,
$R$ is the areal radius, $M_{\rm B}$ is the baryon mass,  and
$\rho_{\rm c}$ is the central energy density, respectively.}
\label{EOStable}
\end{table}

Different aspects of the EOSs show up if we consider the
$1.35~{\rm M}_\odot$ static stars. Their parameters, calculated
using three EOSs are given in Table \ref{EOStable}. Let us remind
that a NS configuration is a {\it functional} of the EOS at the
densities $\rho$ smaller than the central density $\rho_{\rm c}$.
We see that for the GNH3 EOS central density is barely higher than
the hyperon threshold $\rho_{\rm H}$, and therefore the EOS {\it
inside} NS is very stiff, actually the stiffest of all three. 
The largest  stiffness of the GNH3 EOS in the interior of a $1.35~
{\rm M}_\odot$ NS can be clearly recognized by looking at 
Fig. 1  and Fig. 3. Looking at Fig. 1, we see that 
this EOS has the highest $P$ at any 
 $\rho<\rho_{\rm c}$. Additional information is contained in 
Fig. 3: this EOS has particularly large $\gamma$  for $\rho\sim 
\rho_0$, i.e., in the outer layers of NS core, which are therefore 
quite ``inflated'' in comparison with those in the other two 
NS models. These  features are  
responsible for particularly large $R$ for the GNH3 EOS.
The
GNH3  configuration has definitely the largest $R$ and therefore the 
smallest compaction parameter $M/R$. The BPAL12 EOS is clearly the
softest, with $R$ smaller by 27\% than for the GNH3 EOS. Final,
the APR EOS, which is moderately stiff {\it in this mass range},
yields $R$ which is only 8\% larger than for the BPAL12 model.

\begin{figure}[h]
\centering
\resizebox{3.25in}{!}{\includegraphics[clip]{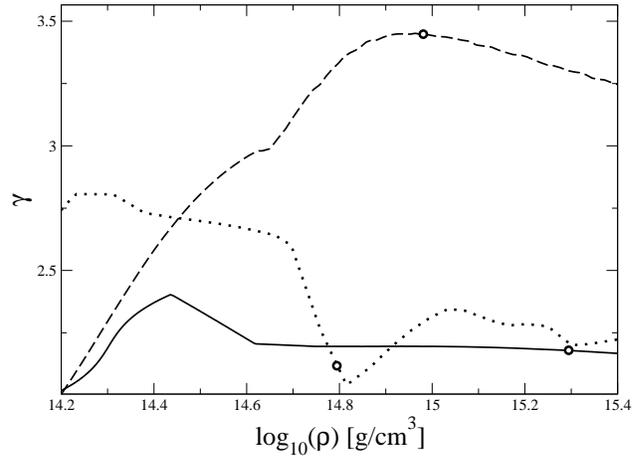}}
\caption{The adiabatic index $\gamma$ against the energy density
$\rho$ for the EOSs used in our computations:
APR (dashed line), GNH3 (dotted line),
and  BPAL12 (solid line). The empty circles correspond to the
central density of a non-rotating stellar model with a
gravitational mass equal to $1.35~M_{\odot}$.}
\label{gamma}
\end{figure}

The differences in stiffness reflect the characteristics of the
nuclear model underlying  each of the EOSs of the NS core. Using
the density dependent adiabatic index $\gamma$ it is particularly
easy to visualize these differences, see Fig.~\ref{gamma}. The
strong drop in $\gamma$ above $\rho_{\rm H}$ reflects the hyperon
softening in the GNH3 EOS. The values of $\gamma$ increasing up to
the maximum at $\rho=\rho_{\rm c}$ tell us about the stiffening of
the APR EOS at $\rho\sim 2\rho_0$, in contrast to the behavior of
the GNH3 EOS which softens close to this density. Finally, the
BPAL12 EOS remains close to a polytrope with $\gamma\simeq 2.2$,
except for a small region around $\rho_0$.

%
\section{Numerical method}
\label{method}
\subsection{Numerical code for close binary systems}

The present computations of close binary neutron star systems rely
on the assumption of quasiequilibrium state (helical Killing
vector approximation), with irrotational flow of the fluid, and
conformally flat spatial 3-metric (Isenberg-Wilson-Mathews
approximation).  We construct the quasiequilibrium sequences of
binary neutron stars described by the realistic EOSs using a
numerical code which solves the five coupled, nonlinear, elliptic
equations for the gravitational field, supplemented by an elliptic
equation for the velocity potential of irrotational flows.  The
code has already been used successfully for calculating the final
phase of inspiral of binary neutron stars described by the
polytropic equation of state (\cite{TanigGB01}, \cite{TanigG02a},
2002b and 2003). This code is built upon the
C++ library {\sc Lorene} ({\tt http://www.lorene.obspm.fr}) and
can be downloaded freely from {\sc Lorene} CVS repository, as {\tt
Lorene/Codes/Bin$\_$star/coal.C}.  The complete description of the
resulting general relativistic equations, the whole algorithm, as
well numerous tests of the code can be found in (\cite{BNSmet01}).
Additional tests have been presented in Sect.~III of
(\cite{BNSvar03}).

The numerical technique relies on a multi-domain spectral method
with surface-fitted coordinates.  We have
used one domain for each star and 3 (resp. 4) domains for the space
around them for a large (resp. small) separation. In each domain,
the number of collocation points of the spectral method is chosen
to be $N_r \times N_{\theta} \times N_{\varphi} = 25 \times 17
\times 16$, where $N_r$, $N_{\theta}$, and $N_{\varphi}$ denote
the number of points in the radial, polar, and azimuthal directions
respectively.  The accuracy of the computed relativistic models
was estimated using a relativistic generalization of the the
Virial Theorem (\cite{FriedUS02}; see also Sec.~III.A of
\cite{BNSvar03}). The virial relative error was a few times $10^{-5}$.

\subsection{The velocity potential of irrotational flows}

Let us briefly discuss a technical difference between the computations of the
 realistic EOS irrotational binary NS and the polytropic case
 considered by
Bonazzola et al. 1999, Maronetti et al. 1999, Uryu \& Eriguchi 2000,
Uryu et al. 2000, Gourgoulhon et al. 2001, 
and Taniguchi \& Gourgoulhon
2002b, 2003. The difference stems from the
fact that the realistic EOS presented in Sec.~\ref{EOSdescr} are
given in tabulated form, and a certain thermodynamic coefficient
$\zeta$, (required in the computation of the velocity potential of
the irrotational fluid flow, cf. Sect. II of \cite{BNSmet01}) is
not given explicitly by the tabulated EOS.

As already discussed in Sect.\ 2, we adopt the approximation of
the perfect fluid for the form of the stress-energy tensor, and we
represent  the NS matter by  a zero-temperature EOS. In view of
this, we can use the Gibbs-Duhem identity:
%
\begin{equation}
\frac{{\rm d} p}{e+p}=\frac{1}{h}{\rm d} h
\label{GDrelation}
\end{equation}
%
where $e=\rho c^2$ is the proper energy density of the fluid, $p$ is the
pressure, $h=(e+p)/(m_{\rm b}n)$ is the specific enthalpy, with $n$
denoting the fluid baryon number density, and $m_{\rm b}$ the
mean baryon mass.

Following  \cite{BNSmet01}, we write the equation for the
coefficient $\zeta$ as
%
\begin{equation}
\zeta = \frac{{\rm d(ln}H)}{{\rm d(ln}n)} = \frac{{\rm
d(ln}H)}{{\rm d(ln}p)}\cdot \frac{{\rm d(ln}p)}{{\rm d(ln}n)} =
\frac{{\rm d(ln}H)}{{\rm d(ln}p)}\cdot\gamma~, \label{zeta}
\end{equation}
%
where $\gamma$ is the adiabatic index. The quantity $H:={\rm ln}h$
appears in the first integral of fluid motion and denotes the
log-enthalpy of the fluid, and the value of ${\rm d(ln}H)/{\rm
d(ln}p)$ can easily be evaluated by means of
Eq.~(\ref{GDrelation}) which, as a strict thermodynamic relation,
is {\it exact}.

In practice, we used two methods to get the adiabatic index
 $\gamma=(n/p){\rm d}p/{\rm d}n$. In the first method we used an
 analytical formula for the adiabatic index being approximation of
 tabulated values of $\gamma$. In the second method the adiabatic
 index was obtained directly from a tabulated EOS by taking the
 derivative of the second order polynomial which goes through three
 consecutive $(p,~n)$ EOS points. The value of the ${\rm d}p/{\rm d}n$
 multiplied by the $n/p$ ratio is evaluated in the middle point, and
 the resulting discrete values of $\gamma$ are ready to be
 interpolated, similar to the other quantities from the tabulated EOS.

The second method was proved to be very robust, and acts as a
consistency check; it was tested with EOS for which the precise
values of the adiabatic index were obtained from the microscopic
considerations, namely for the SLy EOS (\cite{SLy00}).

\section{Numerical results}
\label{results}

\subsection{Quasiequilibrium sequences of neutron stars described by 
realistic EOS}

\label{s:num_seq}
\begin{figure*}[t]
\centering
\includegraphics[angle=-90,clip,width=0.49\textwidth]{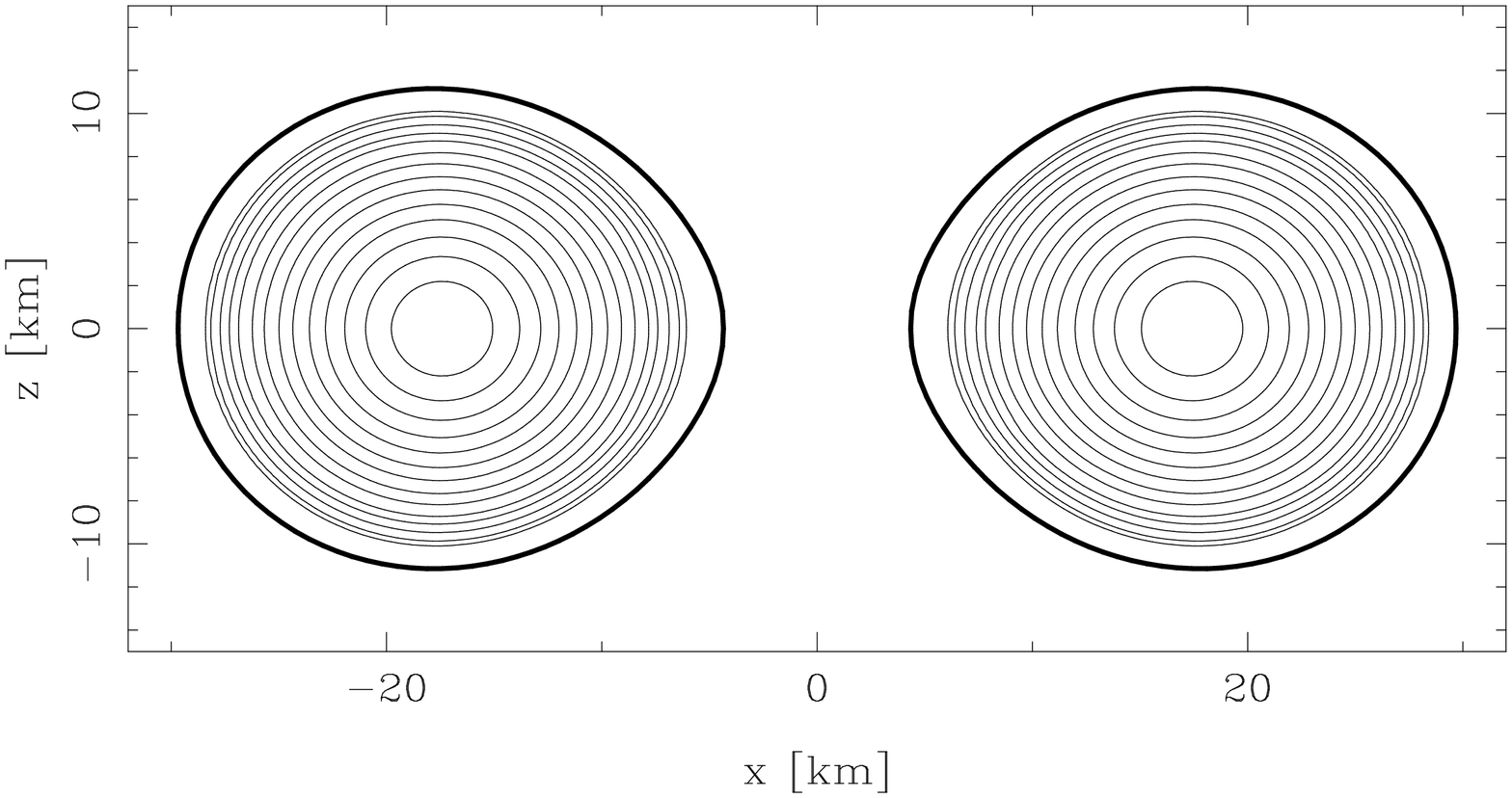}
\includegraphics[angle=-90,clip,width=0.49\textwidth]{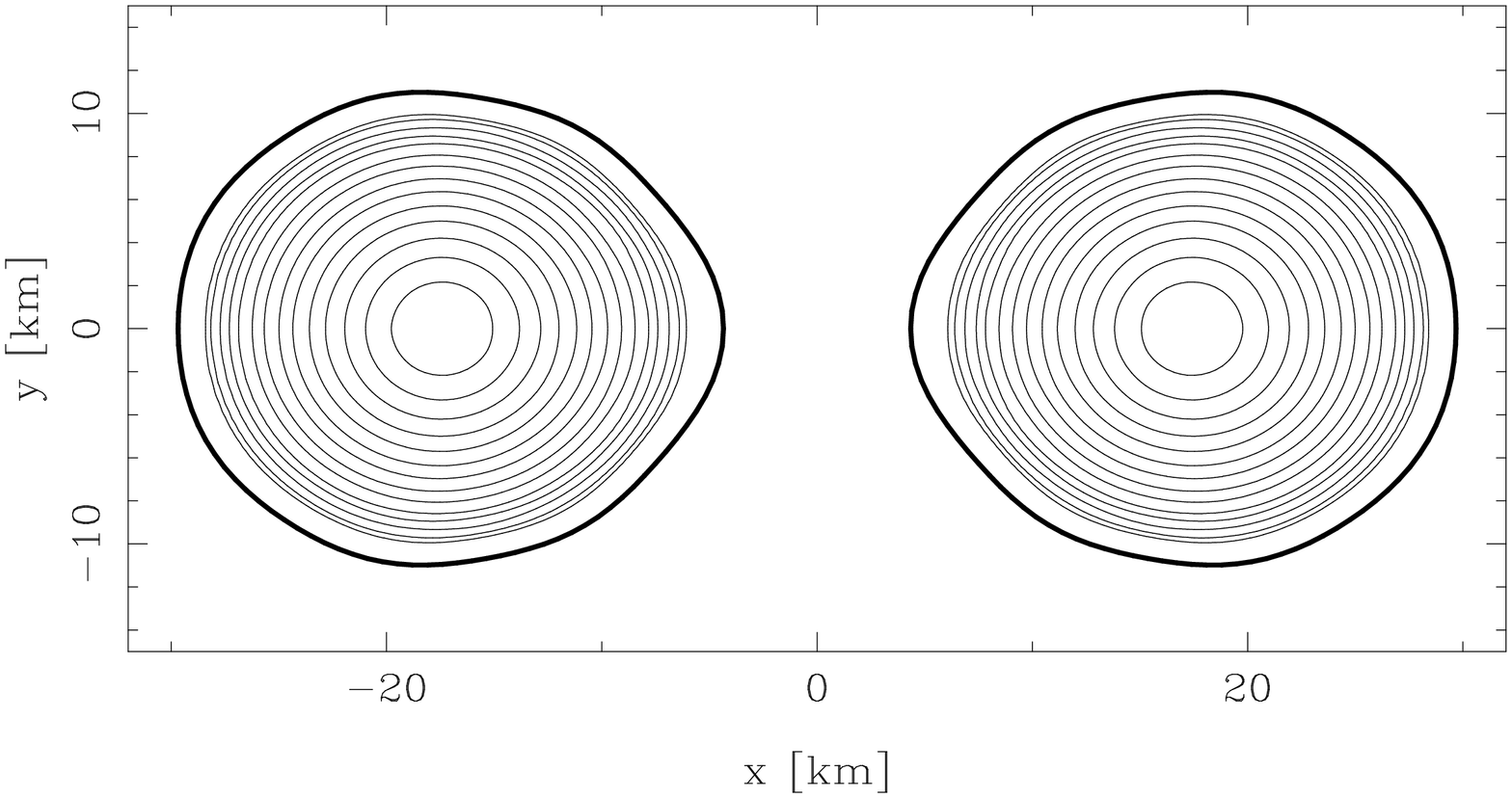}
\includegraphics[angle=-90,clip,width=0.49\textwidth]{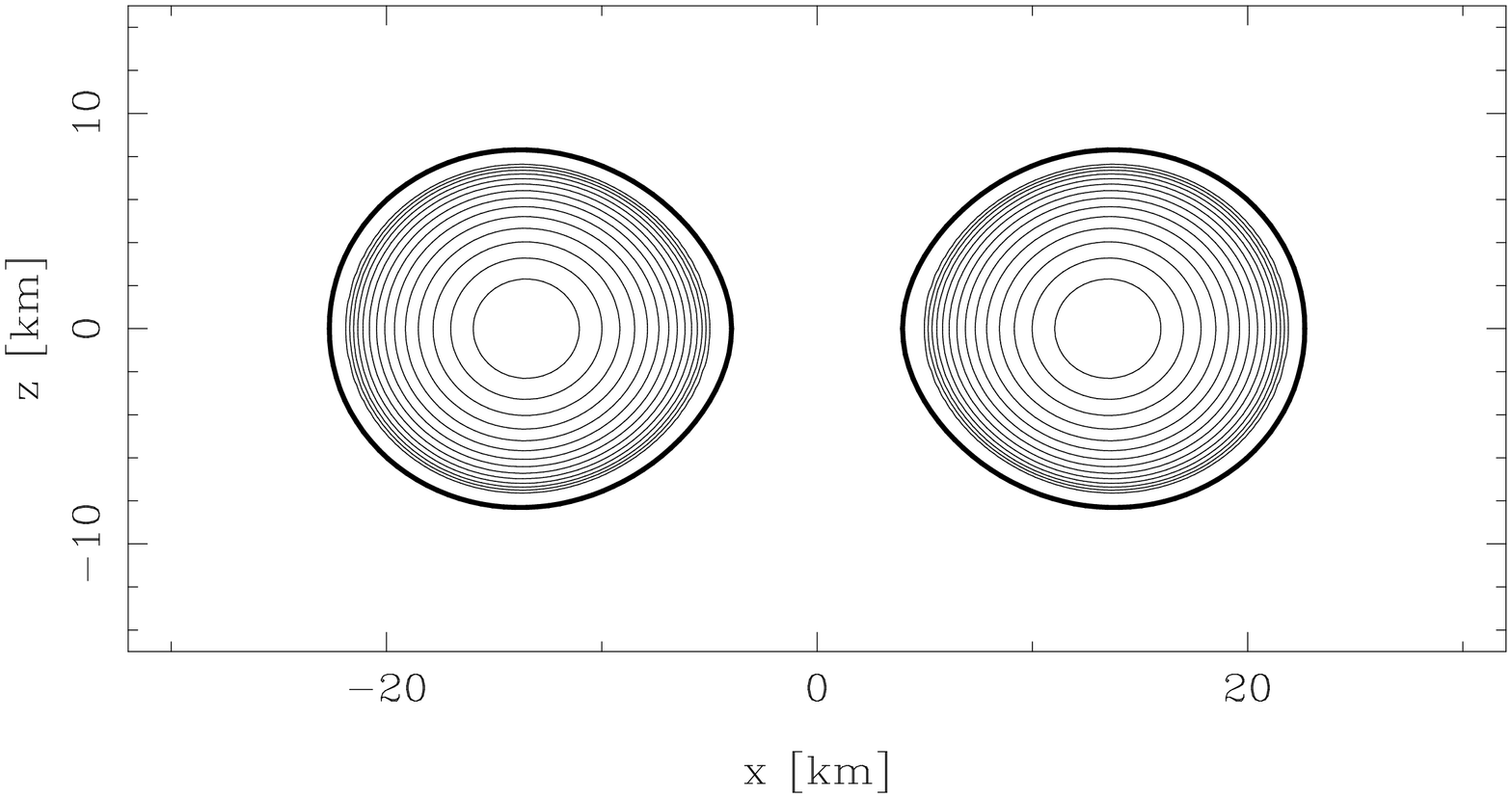}
\includegraphics[angle=-90,clip,width=0.49\textwidth]{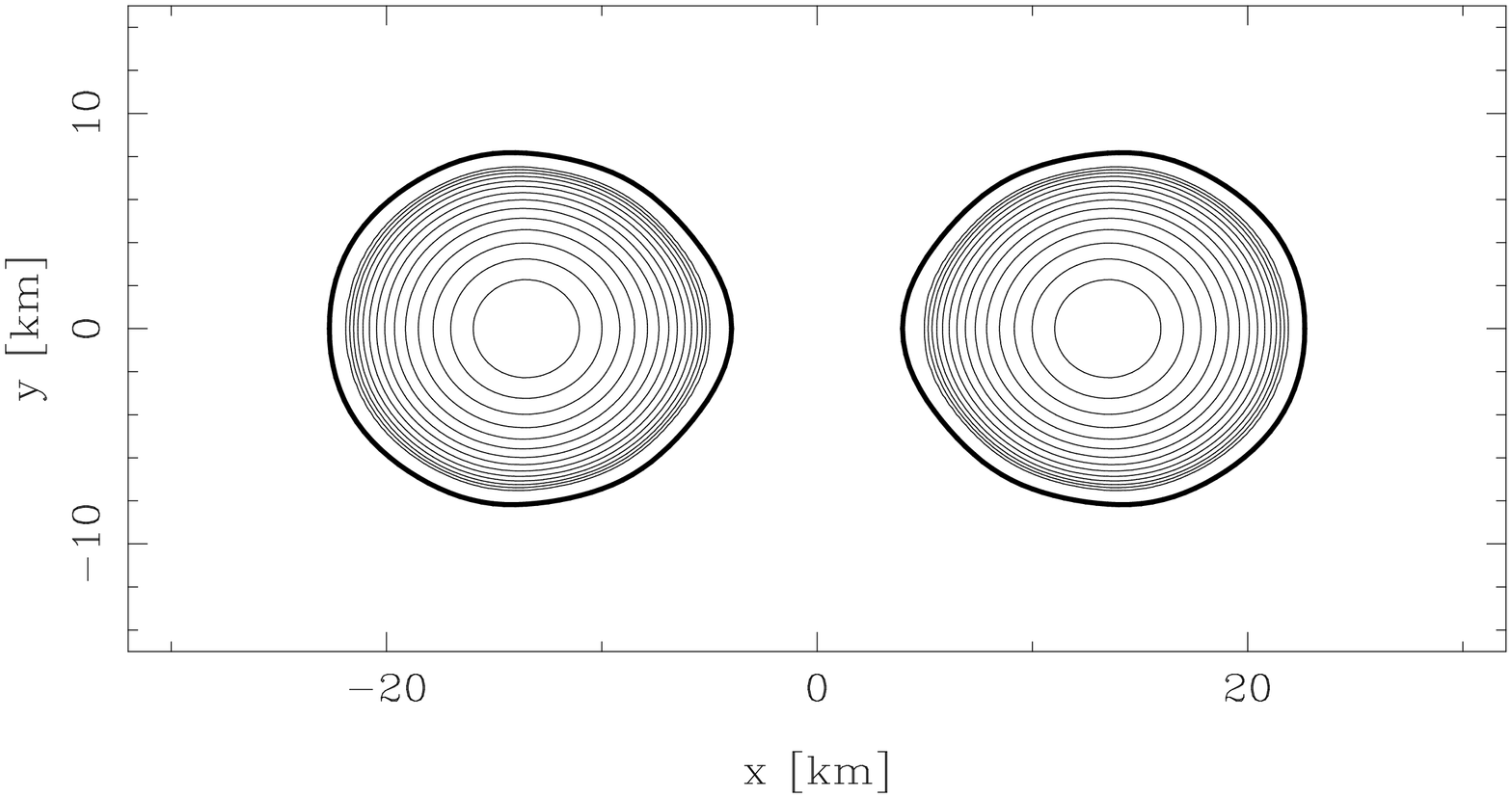}
\includegraphics[angle=-90,clip,width=0.49\textwidth]{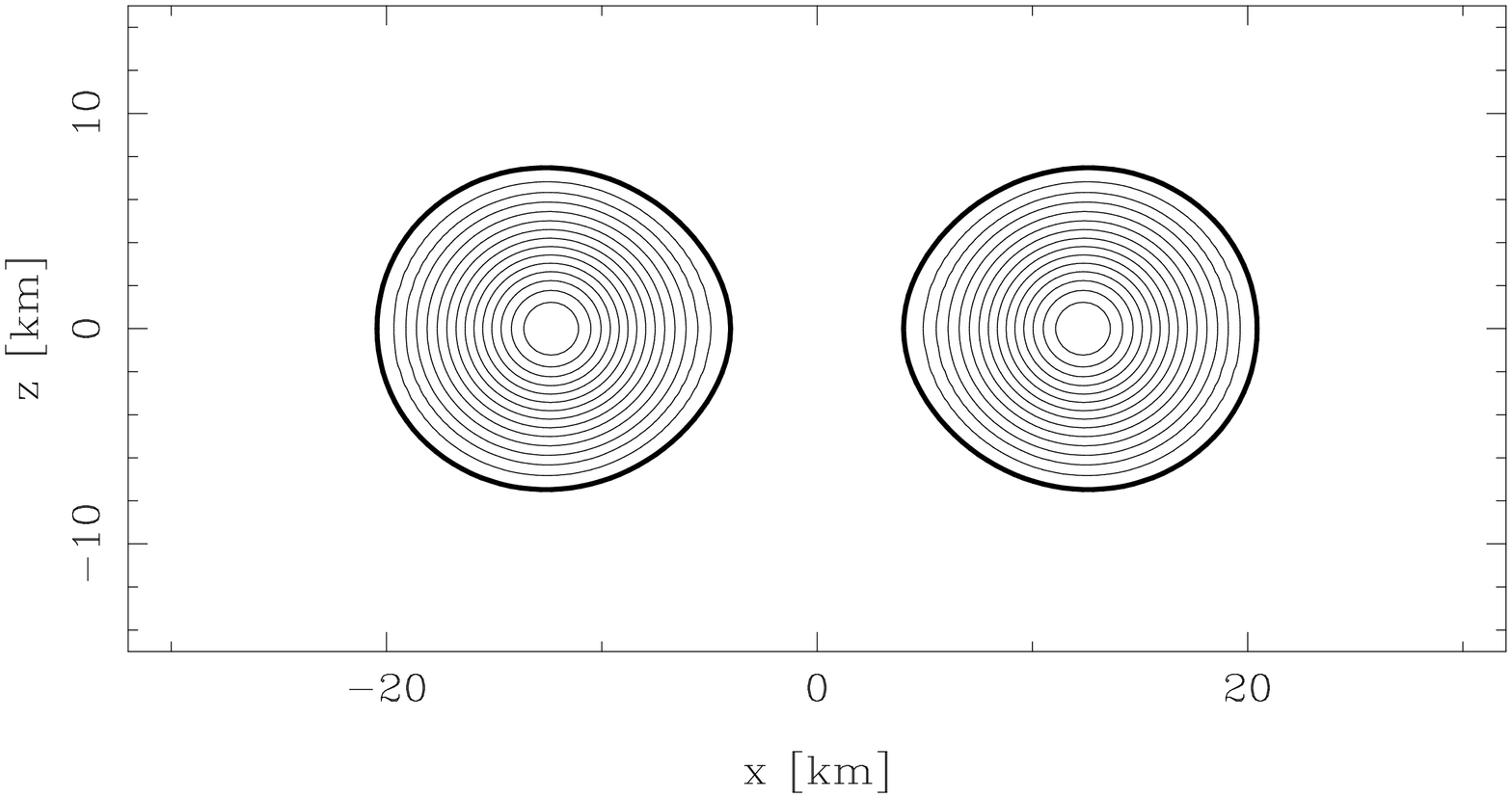}
\includegraphics[angle=-90,clip,width=0.49\textwidth]{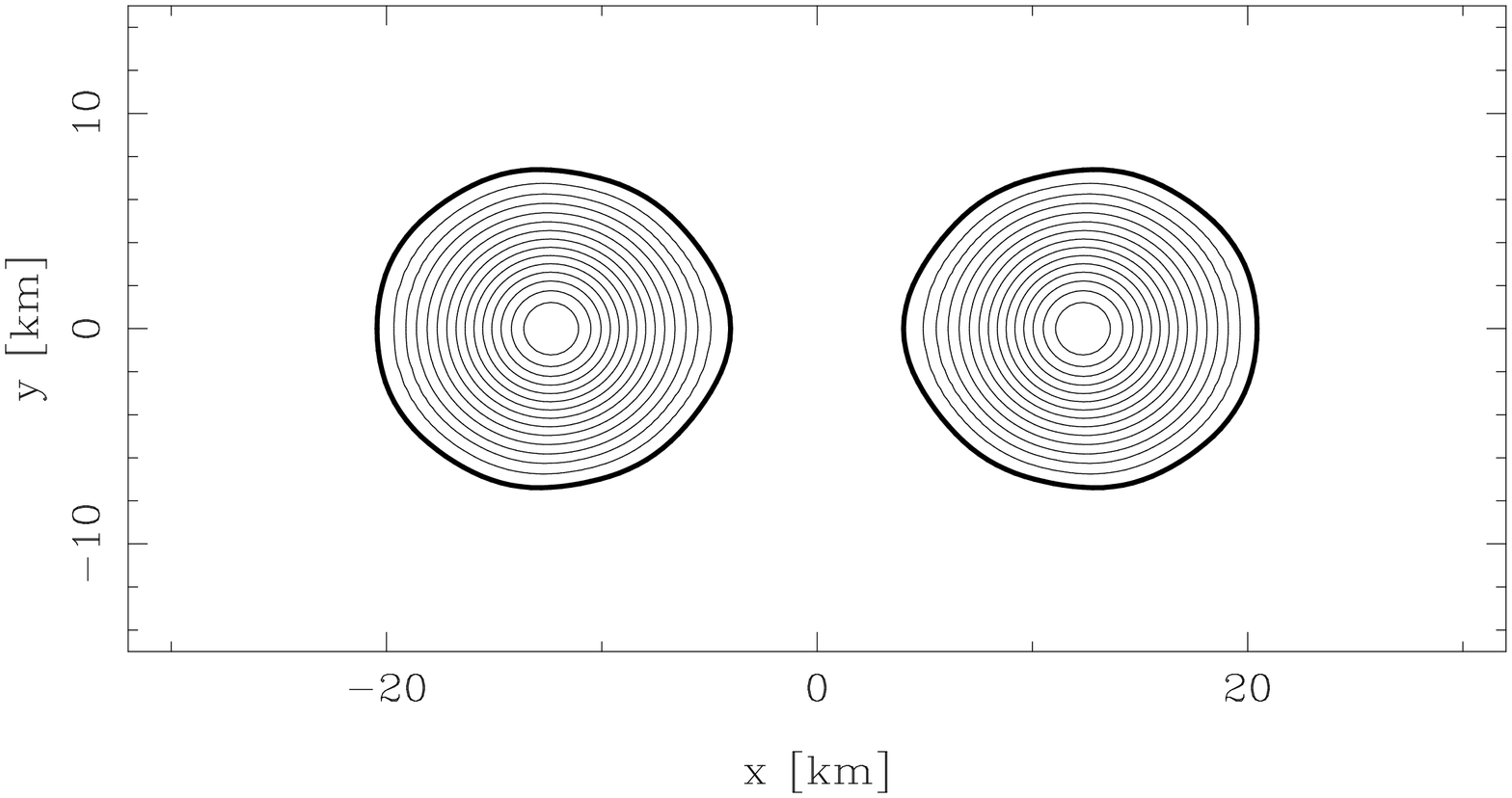}
\caption{Baryon number density isocontours in the coordinate plane
$y=0$ (containing the rotation axis) (left panels) and in the
coordinate plane $z=0$ (orbital plane) (right panels), for
configurations close to the mass-shedding limit. 
The upper (resp. middle, lower) panels correspond to the GNH3
(resp. APR, BPAL12) EOS. The thick solid lines denote
stellar surfaces.}
 \label{bin_prof}
\end{figure*}

In this section we present the numerical results for evolutionary
sequences of close neutron star binaries described by the three
realistic EOS introduced in Sect.~\ref{EOSdescr}. By {\em
evolutionary sequence}, we mean a sequence of quasiequilibrium
configurations of decreasing separation $d$ and with constant
baryon mass $M_{\rm B}$. Such a sequence is expected to
approximate pretty well the true evolution of binary neutron
stars, which is entirely driven by the reaction to gravitational
radiation and hence occur at fixed baryon number and fluid
circulation (zero in the irrotational case considered here). We
calculated evolutionary sequences of binary system composed of two
identical neutron stars (equal mass system), with the  baryon mass $M_{\rm B}$
corresponding to the  gravitational mass $M:=M_1=M_2=1.35 \msol$
for a static isolated star
 (see Table~\ref{EOStable} for the values of $M_{\rm B}$).
We have selected the gravitational mass $M=1.35 \msol$ for two reasons:
(i) it agrees with ``average NS mass'' obtained from
observations of radio pulsars in binary system and (ii) it allows
us to compare our results with calculations made by other authors
for polytropic models (\cite{Faber02}).

For the benefit on the present discussion, let us
recall the definition of the ADM mass
of the system, which measures the total energy
content of a slice $t={\rm const}$
of spacetime (hypersurface $\Sigma_t$):
\begin{equation}
    M_{\rm ADM} := \frac{1}{16\pi} \oint_\infty
    \left[ {\cal D}^j \gamma_{ij} - {\cal D}_i
    \left( f^{kl} \gamma_{kl} \right) \right] dS^i ,
\label{e:M_ADM_def}
\end{equation}
where $\gamma_{ij}$ is the metric induced by the spacetime metric
on $\Sigma_t$, $f_{ij}$ is a flat metric on $\Sigma_t$ (the
condition of asymptotic flatness being $\gamma_{ij} \rightarrow
f_{ij}$), ${\cal D}_i$ is the covariant derivative associated with
$f_{ij}$, and the integral is to be taken on a sphere at spatial
infinity. In the case considered here of a conformally flat
spatial metric, $\gamma_{ij} = \Psi^4 f_{ij}$, the surface
integral (\ref{e:M_ADM_def}) can be converted into a volume
integral over  the whole hypersurface $\Sigma_t$:
\begin{equation}
    M_{\rm ADM} := \int_{\Sigma_t}
                    \Psi^5 \left( E +\frac{1}{16\pi} K_{ij} K^{ij} \right)
                                d^3 x ,        \label{e:M_ADM_vol}
\end{equation}
where $E$ is the total energy density of matter with respect to
the observer whose 4-velocity is normal to $\Sigma_t$ and $K_{ij}$
denotes the extrinsic curvature tensor of $\Sigma_t$.

At infinite separation, the ADM mass of the system, 
Eqs.~(\ref{e:M_ADM_def}) and
(\ref{e:M_ADM_vol}), 
tends toward the sum of the gravitational masses of isolated static stars,
and will be denoted by $M_\infty$:
\begin{equation}
    \lim_{d\rightarrow\infty} M_{\rm ADM} = M_\infty := M_1+M_2
        = 2.7\msol .
\end{equation}
We then define the {\em orbital binding energy} by
\begin{equation}
E_{\rm bind} := M_{\rm ADM} - M_\infty .
\end{equation}
The variation of $E_{\rm bind}$ along an evolutionary sequence
corresponds to the loss of energy via gravitational radiation.
Gravitational waves are emitted mostly at twice
the orbital frequency: $\fgw=2f$.

\subsection{Innermost stable circular orbit}

Each evolutionary sequence terminates by a mass-shedding point,
which marks the end of existence of quasiequilibrium configurations.
The shape of the stars close to this limit is presented in Fig.~\ref{bin_prof}.
The mass-shedding is revealed by the formation of a cusp
at the stellar surface in the direction of the companion
(Roche lobe overflow). This cusp
is marginally visible in Fig.~\ref{bin_prof}. 

A turning point of $E_{\rm bind}$ along an evolutionary sequence
would locate an orbital instability (\cite{FriedUS02}).
This instability originates both from relativistic effects (the well-known
$r=6M$ last stable orbit of Schwarzschild metric) and hydrodynamical
effects (for instance, such an instability exists for sufficiently
stiff EOS in Newtonian regime, see e.g. \cite{TanigGB01} and 
references therein). It is secular for
synchronized systems and dynamical for irrotational ones, as those
considered here. 
Thus the quasiequilibrium inspiral of binary neutron stars 
can terminate by either the orbital instability (turning point
of $E_{\rm bind}$) or the mass-shedding limit.
In the latter case, one may wonder whether there might exist 
equilibrium configurations beyond the mass-shedding limit, 
i.e. dumb-bell like configurations (see e.g. \cite{EriguH85}).
However dynamical calculations for $\gamma = 2$ polytrope
have shown that the time to coalescence was shorter than one 
orbital period for configurations at the mass-shedding limit
(\cite{ShibaU01},\cite{MarronDSB04}). Therefore we may safely
define the end of quasiequilibrium inspiral by the mass-shedding
limit in the case where no turning point of $E_{\rm bind}$ 
is found along the sequences
(which is actually the present case, as we shall discuss below).

\begin{figure}[h]
\centering
\resizebox{3.25in}{!}{\includegraphics[angle=-90]{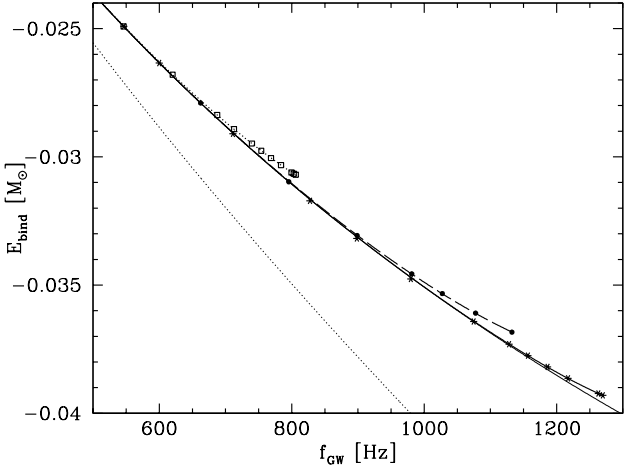}}
\caption{Orbital binding energy $E_{\rm bind} = M_{\rm ADM} -
M_\infty$ of the binary system versus frequency of gravitational
waves (twice the orbital frequency) along three irrotational
quasiequilibrium sequences. The lines (dotted - GNH3,  dashed -
APR, solid - BPAL12) were plotted using the fit described in the
text, whereas the points represent actual data. Thin solid line
touching the bottom-right corner shows the 3rd post-Newtonian
approximation for point masses derived by \cite{Blanc02}. The
lower thin dotted curve corresponds to the Newtonian limit for
point masses.}
\label{fig:Mfgw}
\end{figure}

The variation of the orbital binding energy along evolutionary
sequences is presented in Fig.~\ref{fig:Mfgw}, where the points
correspond to the equilibrium configurations of binary system
calculated using numerical method (see Sect.~\ref{method}) and the
lines present our best fit described in detail in
Sect.~\ref{sec:fit}. In the scale of Fig.~\ref{fig:Mfgw} there is
no visible difference between our numerical results and the fit,
i.e. fitting curves pass through the points. We plotted also the
binding energy obtained in the framework of Newtonian theory and
the 3PN post-Newtonian approximation for point masses
(\cite{Blanc02}). Figure~\ref{fig:Mfgw} shows clearly that no
turning point of $E_{\rm bind}$ occurs along evolutionary sequences.
Hence there is no orbital instability prior to the mass-shedding
limit. We conclude that the 
{\em innermost stable circular orbit (ISCO)} of the computed 
configurations are given by the mass-shedding limit. 

\begin{figure}[h]
\centering
\resizebox{3.25in}{!}{\includegraphics[angle=-90]{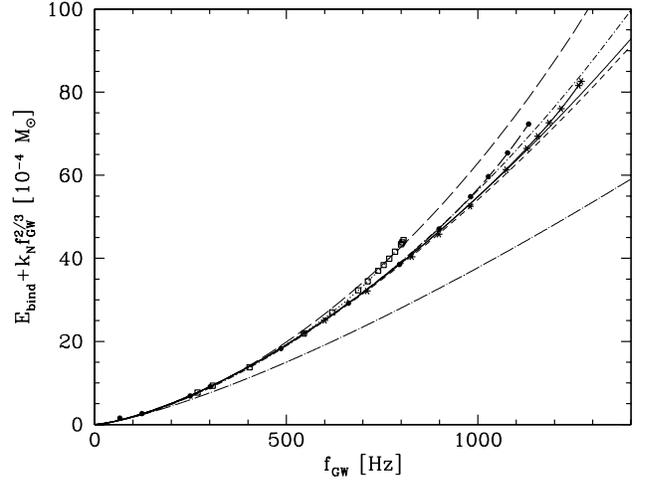}}
\caption{ Orbital
binding energy of the binary system minus the (point mass)
Newtonian term $-k_{N} f^{2/3}$ versus frequency of gravitational
waves (twice the orbital frequency) along three irrotational
quasiequilibrium sequences. Thin solid line shows the 3rd
post-Newtonian approximation for point masses by \cite{Blanc02}.
Slightly below there is the 2PN line (thin, dashed) and the 1PN
one (thin, long dash-dotted). For comparison post-Newtonian
approximation by \cite{DamouJS00} - 3PN (dashed-dotted thin line)
and 2PN (thin, long dashed) are also plotted.
 }
\label{fig:admrel}
\end{figure}

To compare different approaches to the relativistic description of binary
systems we present in Fig.~\ref{fig:admrel} the orbital binding
energy after subtraction of the Newtonian term $\propto \fgw^{3/2}$.
This figure shows thus the effects of general relativity and of
finite sizes (hydrodynamics).
One can see that binary neutron star systems are quite far from
Newtonian and 1PN systems. On the contrary, 2PN and 3PN results
by \cite{Blanc02} and 3PN results by \cite{DamouJS00}
are very close to our results for a wide range of frequencies.
Note that the difference between 2PN and 3PN approximations are much larger
in \cite{DamouJS00} treatment (Effective One Body method)
than in that of \cite{Blanc02} (standard post-Newtonian expansion).

\subsection{Analytical fits to the numerical results}
\label{sec:fit}

Following \cite{Faber02}, we have performed some polynomial fits to
 each of the computed sequences.
This is a very important step  in order to obtain the first
derivative of the functions required for the energy spectrum of
gravitational waves (see below). We used two different approaches
to approximate our numerical results. The first one, similar to
that presented by \cite{Faber02}, is based on quadratic
approximation of numerical results. First of all we decided to
make a fit only to the difference between obtained exact results
for the ADM mass $M_{\rm ADM}$ [cf. Eq.~(\ref{e:M_ADM_def})] and
the prediction of the Newtonian theory - i.e. we made a fit to the
function $ E_{\rm bind} + k_{\rm N}\fgw^{2/3}$, where the second
term corresponds to the Newtonian point-like mass behaviour with
$k_{\rm N}=(G\pi/4)^{2/3}M^{5/3} = 4.06\, 10^{-4}~M_{\odot}{\rm
Hz}^{-2/3}$. We found it sufficient to fit the numerical results
by a second order polynomial without any linear term:
\begin{equation}
\label{fit}
E_{\rm bind} = -k_{\rm N}\fgw^{2/3}+k_{\rm 2}\fgw^2, 
\end{equation}
i.e. contrary to \cite{Faber02} we assume $k_1=0$. The best-fit
coefficients $k_{\rm 2}$ are collected in Table \ref{fittable}. It should be
mentioned that our aim was to obtain accurate formula for
gravitational radiation in the region where it is effectively
emitted and therefore we have performed the fitting procedure for
frequencies larger than 500-600 Hz. This approximation works quite
well for rather large frequencies (i.e. small distance between
stars) and there is no need to introduce the linear term, as 
done by Faber et al (2002). The advantage of our
quadratic formula is its simplicity and good accuracy in the
region corresponding to the effective emission of gravitational
waves.

However it is possible to find much better approximation of the
numerical results if one takes into account high order
post-Newtonian approximation for the binding energy of point-mass
systems. The 3PN formula as obtained by \cite{Blanc02} from the
standard\footnote{i.e. non-resummed, in contrast to the so-called
Effective One Body approach of \cite{DamouJS00}} post-Newtonian
expansion reads
\begin{eqnarray}
\frac{E_{\rm bind}^{\rm 3PN}}{M_\infty} &=&
-{1\over 8} \, {\Omega_*}^{2/3}
+ {37\over 384}\, {\Omega_*}^{4/3} + {1069\over 3072} \, {\Omega_*}^{2}
                    \nonumber \\
 &&+ {5\over3072} \left({41} \pi^2 -{285473\over 864}\right)
 \, {\Omega_*}^{8/3} , \label{pn3luc}
\end{eqnarray}
where $\Omega_*$ is the orbital angular frequency expressed in
geometrized units:
$$
\Omega_* :=2\pi {GM_{\infty}\over c^3} f = 2\pi {GM\over c^3} \fgw.
$$
The terms in ${\Omega_*}^{2/3}$, ${\Omega_*}^{4/3}$, ${\Omega_*}^{2}$ and
${\Omega_*}^{8/3}$ in Eq.~(\ref{pn3luc}) are respectively
the Newtonian, 1PN, 2PN and 3PN term.

\begin{figure}[h]
\centering
\resizebox{3.25in}{!}{\includegraphics[angle=-90]{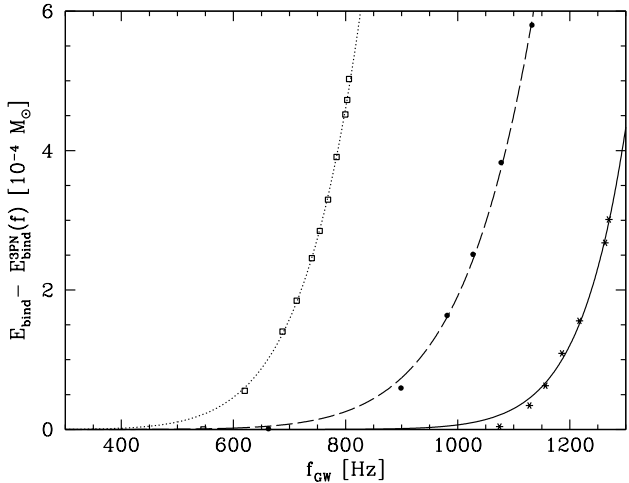}}
\caption{Difference $E_{\rm bind} - E_{\rm bind}^{\rm 3PN}$ between
the binding energy of irrotational binary neutron stars build upon realistic
equation of state and the binding energy of binary point masses in the
3PN approximation of \cite{Blanc02}.
The dots correspond to numerical results and the lines to polynomial
fits to them (see text for details). }
\label{fig:aproxl}
\end{figure}

In Fig.~\ref{fig:aproxl} we present the difference between our
numerical results and the 3PN approximation given by
Eq.~(\ref{pn3luc}). Formula~(\ref{pn3luc}) approximates very well
the behavior of binary system of realistic neutron stars for a very
large range of binary period (notice the scale of the y-axis of
Fig.~\ref{fig:aproxl} !). From Fig.~~\ref{fig:aproxl}, we can
define frequencies $f_{\rm npm}$ as those frequencies at which the
deviation from point-mass behavior becomes important. The values
of these frequencies for each of the three considered EOSs are
given in Table \ref{ftable}. Of course,  they are not precise
numbers and should be treated as having a $\sim \pm 50\,{\rm Hz}$
uncertainty. We can approximate the results presented in Fig.
\ref{fig:aproxl} by the power-law  dependence on frequency $\fgw$:
\begin{equation}
\label{aprl}
 E_{\rm bind} - E_{\rm bind}^{\rm 3PN} =
    a\,\left({\fgw\over 1000 {\rm Hz}}\right)^n
\end{equation}
Because of the steep
character of the function $E_{\rm bind} - E_{\rm bind}^{\rm 3PN}$
seen in Fig.~\ref{fig:aproxl},  the power $n$ is quite large.
The values are listed in Table~\ref{fittable}.
It should be stressed  that we assume the integer number of the 
power $n$, although this is in principle not obvious from the 
theoretical point of view. Of course the more careful treatment of this
approximation is possible (eg. more terms in the expansion) but from
Fig. \ref{fig:aproxl} it is clear that there is no need to do it and
the leading term of high power is sufficient. 

One can draw an important conclusion from the presented results and
their comparison with relativistic approximations for point masses
in binary system. We can expect that taking into account next
orders in post-Newtonian approximation does not change the energy
by an amount larger than the difference between 2PN and 3PN
models. As a consequence the large deviation of our numerical
results from the 3PN approximation is caused by the effect of a
finite size of the star (e.g. tidal forces). The very high power $n$
in relation (\ref{aprl}) indicates that, even 
for small departures from point mass approximation, 
high-order tidal effects are very important,
and dominate the relation $ E_{\rm bind} (\fgw)$.
Indeed the lowest order tidal term is known to be $n=4$ (\cite{LaiRS94})
and the values obtained here are well above this. 

\begin{table}[h]
\begin{center}
\begin{tabular}{c|c|c|c}
\multicolumn{1}{c}{EOS} &
\multicolumn{1}{|c|}{$k_{\rm 2}~[10^{-9} \msol\, {\rm s}^2]$}&
\multicolumn{1}{|c|}{$a~[\msol]$} &
\multicolumn{1}{|c}{$n$} \\
\hline\hline
GNH3 & 6.23 &$2.745\,10^{-3}$ &8 \\
\hline
APR  & 5.38 &$1.912\,10^{-4}$ &9 \\
\hline
BPAL12 & 4.59 &$6.519\,10^{-6}$ &16 \\
\end{tabular}
\end{center}
\caption{Parameters of polynomial fits (\ref{fit}) and (\ref{aprl}).}
\label{fittable}
\end{table}

\subsection{Energy spectrum of gravitational waves for different realistic EOS}
\begin{figure}[h]
\centering
\resizebox{3.25in}{!}{\includegraphics[angle=-90]{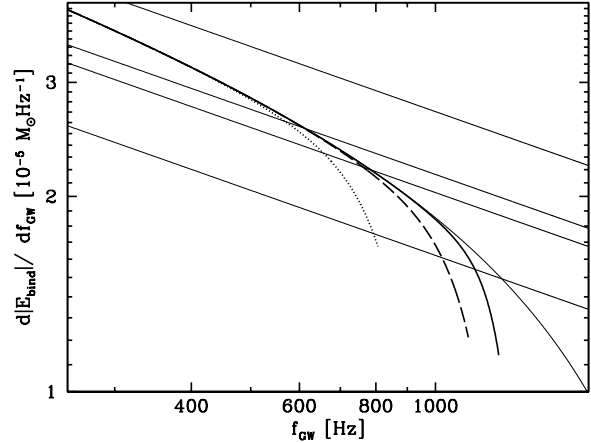}}
\caption{Energy spectrum of GW waves emitted by the  binary neutron star system
versus frequency of gravitational waves (twice the orbital frequency)
along three irrotational quasiequilibrium sequences.
The straight lines correspond to the Newtonian dependence of energy
multiplied by 1, 0.8, 0.75 and 0.6. }
\label{fig:dMdf}
\end{figure}

We computed the energy spectrum of
gravitational waves obtained as the first derivatives of the
fitted functions [Eq. (\ref{aprl})]. The relation between 
${\rm d}E_{\rm bind}/{\rm d}f$ and the GW frequency $f_{\rm GW}$
is presented on Fig.~\ref{fig:dMdf}. In this figure we draw also
straight lines corresponding to the Newtonian case $\sim
\fgw^{2/3}$ to find the break frequencies at which the energy
spectrum has dropped by 20\%, 25\%, 40\%. These values are 
important from the point of view of the future detections: they
show the difference between the amplitude of the real signal and 
the Newtonian template which allow to calculate the real wave form
amplitude from the detector noise. We also compare our results 
with the 25\% case from \cite{Faber02}.
It should be mentioned that there is no visible difference between
our models for different EOSs at the break frequency level of 10\%
(the case considered by \cite{Faber02}) and the situation is then
very precisely  described by the 3PN formula.

\begin{table}[h]
\begin{center}
\begin{tabular}{c|c|c|c|c|c}
\multicolumn{1}{c}{EOS} &
\multicolumn{1}{|c|}{$f_{\rm npm}$} &
\multicolumn{1}{|c|}{$f_{\rm 20}$} &
\multicolumn{1}{|c|}{$f_{\rm 25}$}&
\multicolumn{1}{|c|}{$f_{\rm 40}$} &
\multicolumn{1}{|c}{$f_{\rm end}$} \\
\hline\hline
GNH3 &500&567&657&792&806  \\
\hline
APR &700&615&762&1025& 1132  \\
\hline
BPAL12 &1050&615&785&1160&1270  \\
\end{tabular}
\end{center}
\caption{Gravitational wave frequencies (in Hz) computed from 
the calculated data for GNH3, APR and BPAL12 EOSs: 
the $f_{\rm npm}$ denotes the
frequency at which the non-point-mass effects start to be important,
$f_{\rm 20},~f_{\rm 25}$ and $f_{\rm 40}$ are the so-called
break-frequencies (see text), whereas $f_{\rm end}$ is the 
GW frequency on the final orbit.}
\label{ftable}
\end{table}

\subsection{Comparison with polytropic EOS}
\label{cmppoly}
%
 Up to now, all calculations (except those of
\cite{OechsUPT04}) of the hydrodynamical inspiral and merger
phases have been done for the simplified equation of state of
dense matter, for the polytropic EOS, where the dependence between
pressure and baryon density is given by $p=\kappa
{n}^{\gamma}$.  It has been  shown that the results obtained
for these polytropic EOSs  depend mainly  on the compaction
parameter $M/R$. It is therefore interesting to check if the
properties of inspiraling neutron stars described by realistic EOS
can be predicted, in a  good approximation, by studying binaries
with assumed polytropic EOSs. In order to make such a comparison
we construct sequences of binary NSs described by polytropic EOS
parametrized by compaction parameter $M/R$  given in the Table~\ref{EOStable}
for each of three realistic equation of states. For a given
$\gamma$, we calculate  the value of the $\kappa$ coefficient
which yields the same $R$ at  $M = 1.35 M_{\odot}$ as that
predicted by a selected realistic EOS used in the present paper.
The values of $\gamma$, the pressure coefficient
 $\kappa$, and  the baryon masses of polytropic static isolated
 NSs  of $M=1.35M_\odot$ are  presented in Table \ref{POLYtable}.

\begin{table}[h]
\begin{center}
\begin{tabular}{c|c|c|c}
\multicolumn{1}{c}{corresponding EOS} &
\multicolumn{1}{|c|}{$\gamma$} &
\multicolumn{1}{c}{$\kappa \  [\hat{\rho}c^2 \hat{n}^{-\gamma}]$ } &
\multicolumn{1}{|c}{$M_{\rm B} \ [M_{\odot}]$} \\
\hline
\hline
GNH3 & 2. & 0.02645 & 1.44802 \\
\hline
& 2. & 0.02037 & 1.47016\\
\noalign{\vskip-5pt}
APR & &  \\
\noalign{\vskip-5pt}
\cline{2-4}
& 2.5 & 0.01094 & 1.49561\\
\hline
BPAL12 & 2. & 0.01908 & 1.47742
\end{tabular}
\end{center}
\caption{Pressure coefficient $\kappa$, adiabatic index $\gamma$  
and baryon mass $M_{\rm B}$ for polytropic NSs having the 
same compactness parameter
and mass ($M=1.35\, M_\odot$) than NSs described by realistic EOSs 
(Table \ref{EOStable}) ($\hat{\rho}:=1.66\times
10^{14}~{\rm g/cm^3}$ and $\hat{n}:=0.1~{\rm fm^{-3}}$).} 
\label{POLYtable}
\end{table}


\begin{figure}
\centering
\resizebox{3.25in}{!}{\includegraphics[angle=-90]{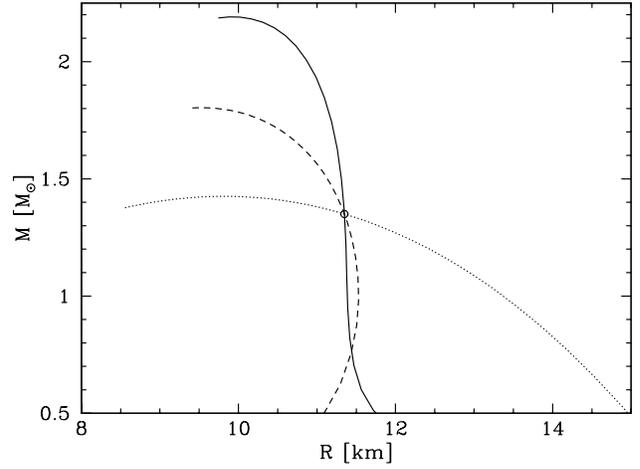}}
\caption{ Gravitational mass versus stellar radius for sequences
of static neutron stars described by \cite{akmal} EOS (solid line) and
polytropic EOS with $\gamma =2$ (dotted line) and $\gamma =2.5$
(dashed line). Configurations with gravitational mass $1.35\msol$
(marked by a circle) described by polytropic EOSs have the same
compaction parameter, $M/R$=0.176, as a neutron star configuration
based on the APR EOS.} \label{fig:MR}
\end{figure}

\begin{figure}
\centering
\resizebox{3.25in}{!}{\includegraphics[angle=-90]{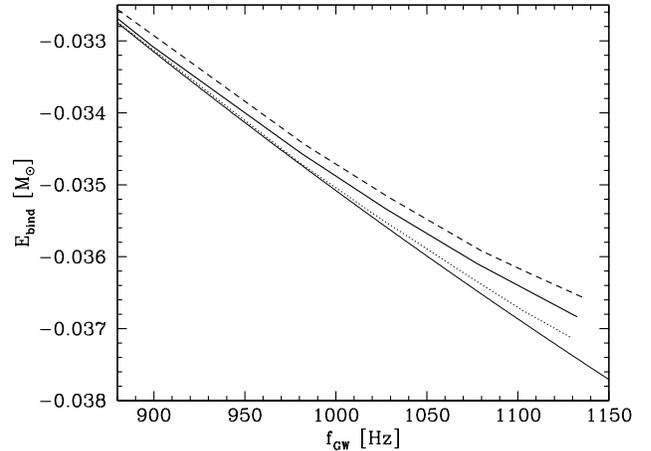}}
\caption{Binding energy of a binary neutron star systems as a
function of gravitational waves frequency for the APR EOS (thick
solid line)  and two corresponding polytropic EOS with $\gamma=2.5$ (dashed line) and   $\gamma=2$ (dotted line). Static isolated NSs have $M/R=0.176$ and $M= 1.35\,\msol$. Thin solid line shows 3PN approximation for point
masses by \cite{Blanc02}, Eq.\ (\ref{pn3luc}).}
\label{fig:poly}
\end{figure}

 In Fig.~\ref{fig:MR} we show the dependence of gravitational mass
versus radius for isolated non-rotating stars based on the APR EOS,
and two different polytropes having $M/R$=0.176 and $M =1.35 M_\odot$
at infinite separation.  Finally in Fig.~\ref{fig:poly} we present
the binding energy versus the frequency of gravitational waves at the
last orbits of the inspiral for the APR EOS and the two corresponding
polytropes. The differences between quasiequilibrium sequences
described by realistic and polytropic sequences are small. For the three
different EOSs (APR and two polytropes) the frequency of gravitational
waves at the last calculated orbit ($f_{\rm end}$) is $\sim 1140 {\rm
Hz}$. Also the binding energy of the system at the mass-shedding point is
close to each other,  between $-0.0372$ and $-0.0366\, M_\odot$.  
However one can see
the differences in the frequencies $f_{\rm npm}$ at which
non-point-mass effects start to be important. The $f_{\rm npm}$ has the
smallest value for the polytrope with $\gamma=2.5$ and the highest for
$\gamma=2$. Although the matter in the stellar interior is stiffer for
the APR EOS ($\gamma \simeq 3$, see Fig.~\ref{gamma}), which is also
clear from Fig.~\ref{fig:MR}, the APR curve lies between $\gamma =
2.5$ and $\gamma = 2$ polytropes in Fig~\ref{fig:poly}.
This is due to the relatively soft equation of
state for the crust for realistic NS models, which makes the response
of the crust to tidal forces different from that of a polytrope
with $\gamma=2.5$ or $\gamma=2$.

We obtained similar results by comparing two other sequences of
realistic EOS - BPAL12 and GNH3 with corresponding polytropic cases.
As it turned out, the irrotational flow is weakly affected by the
changes in the EOS of core, but it is expected that the differences
should be seen in the merger phase. The outer layers of the star
(those with subnuclear densities, i.e., the {\it crust}), which are
properly treated in the present paper have influence on the properties
of binary system at the last stages of inspiral.
However the crucial parameter which determines the energy-frequency
spectrum (energy per frequency) of emitted gravitational energy is
$M/R$.

%
\section{Concluding remarks}
\label{conclusion}
%
We have presented a set of evolutionary sequences 
of binary neutron stars based
on three selected realistic EOSs. These EOSs are based
on modern many-body calculations. Three baryonic EOSs of
neutron-star core have been joined with a recent EOS of neutron-star
crust, and in this way we obtained three different models of
neutron star interior, from the surface to the stellar center. We
restricted to models of neutron star core without exotic phases
(meson condensates, quark matter). In this way, the differences
between the core EOSs reflect the uncertainties in the existing
theories of the interactions in nuclear matter.

In the present paper we considered only those constituents 
of dense matter, which have been studied in laboratory. 
We did not consider here phase transitions to hypothetic exotic
phases of dense matter, which were proposed by many authors, 
but which still remain hypothetical and speculative. Results 
obtained for the  NS-NS binaries with exotic-phase neutron-star 
cores and realistic envelopes will be considered in our future publications. 
Similarly, the case of a binary involving strange quark stars 
built of a self-bound strange quark matter, will also be presented 
in a separate paper. 

We have computed quasiequilibrium sequences of irrotational 
NS-NS binary by keeping constant the baryon mass to a value
which corresponds to individual gravitational masses of $1.35\, M_\odot$ 
at infinite separation. 
For a long time of evolution of the binary system its  binding energy
is very accurately given by the 3PN post-Newtonian formula for point-mass
system. However the departure from this 3PN model at low binary periods has
a quite abrupt character, presumably due to high order tidal effects.
The sequences end at the onset of the mass transfer between the
stars (i.e. when a cusp forms at the surface of the stars).
This point defines the ISCO since no turning points of the binding energy 
has been found along the sequences, which would have revealed some
orbital instability. 
The gravitational wave frequency at the ISCO is 
800~Hz, 1130~Hz and 1230~Hz, for  
the GNH3, APR and BPAL12 EOSs respectively.

In a recent work based on the numerical 
integration of the full set of time dependent Einstein
equations, \cite{MarronDSB04}  have located the 
dynamical ISCO by comparing the time evolution of 
quasiequilibrium initial data at various separations
(see also Fig.~14 of \cite{ShibaU01}).  
This defines the true ISCO, as opposed to the ``quasiequilibrium''
ISCO obtained here.  
For a polytropic EOS with $\gamma=2$ and a compactness
parameter $M/R=0.14$, they obtain the ISCO at an 
orbital frequency which is 15~\% lower than the mass-shedding
frequency of the quasiequilibrium sequence of \cite{BNSgr02}. 
This makes us confident that the values of the gravitational wave frequencies
given above are quite close to those of the end of the true inspiral. 

When comparing our results with those of the recent work of
\cite{OechsUPT04}, one should stress the basic difference in the
EOS of matter at subnuclear densities. Oechslin et al. represented
the EOS of the crust by a polytrope  with the adiabatic index
$\gamma=2.86$. In this way, they made their EOS quasi-continuous at
the crust-core interface, because their nuclear core EOS is very
stiff. However, such an EOS for the crust is very unrealistic,
because the real $\gamma$ can be as low as 0.5 near the neutron
drip point. As the crust EOS is crucial for the last stable orbits
of the NS binary, this implies differences between theirs and our
results, 
 even if for a $M=1.35M_\odot$ model, the compactness
$M/R$ resulting from the GNH3 EOS and their EOS
are very close. In particular, 
\cite{OechsUPT04} have found a turning point in the binding 
energy along their sequences, resulting in a quasiequilibrium ISCO. 
This difference is certainly due to the stiffness of the
polytropic EOS used by them in the outer layers of the star.
Let us recall that for polytropic irrotational binaries a turning point ISCO
exists only if $\gamma > 2.5$ (\cite{UryuSE00}, \cite{BNSvar03}). 

In our calculations we treated the neutron-star matter as an ideal
fluid. In other words, we neglected the elastic shear (in the
crust - if not molten) and viscous (in the crust and in the core)
terms in the matter stress tensor. There terms are believed to be
small, but they lead to specific physical phenomena.  In
particular, the matter flow in NS interior will break the beta
equilibrium between baryons and leptons, and this will imply a
neutrino burst at the last stage of inspiral (see  Haensel 1992).
Moreover, dissipative processes will heat the matter. Both effects
will be studied in future  publications.

All numerical results presented here, including the full 
binary configurations, are available to download from the LORENE
main server {\tt http://www.lorene.obspm.fr/data/}

\begin{acknowledgements}
We are grateful to Luc Blanchet for providing us with the formulae for
the binding energy at various post-Newtonian orders.  We also warmly
thank Koji Uryu for useful discussions and careful reading of the
manuscript.  This work has been funded by the KBN grants 5P03D.017.21,
2P03D.019.24 and PBZ-KBN-054/P03/2001 and has been partially supported
by the Associated European Laboratory Astro-PF (Astrophysics
Poland-France) and  by the ``Bourses de recherche 2004 de la
Ville de Paris''. KT also acknowledges a Grant-in-Aid for Scientific
Research (No. 14-6898) of the Japanese Ministry of Education, Culture,
Sports, Science and Technology.
\end{acknowledgements}

\end{document}